\documentclass{aa}
\usepackage{aabib}
\usepackage{graphicx}
\def\aj{AJ}			
\def\araa{ARA\&A}		
\def\apj{ApJ}			
\def\apjl{ApJ}		
\def\apjs{ApJS}

\def\aap{A\&A}

\def\mnras{MNRAS}

\def\pasp{PASP}		
\def\pasj{PASJ}

\def\nat{Nature}		
\def\iaucirc{IAU~Circ.}

\def\xmm{{\sl XMM-Newton}}
\def\cxo{{\sl Chandra}}
\def\vor{V1647\,Ori}
\newcommand{\oversim}[2]{\lower0.5ex\vbox{\baselineskip=0pt\lineskip=0.2ex
     \ialign{$\mathsurround=0pt #1\hfil##\hfil$\crcr#2\crcr\sim\crcr}}}

\begin{document}

\title{Enhanced X-ray variability from V1647\,Ori,\\
the young star in outburst illuminating McNeil's Nebula}
\titlerunning{Enhanced X-ray variability from V1647\,Ori}
\authorrunning{Grosso et al.}
\author{N.\ Grosso\inst{1}
\and J.H.\ Kastner\inst{2} 
\and H.\ Ozawa\inst{1} 
\and M.\ Richmond\inst{2} 
\and  T.\ Simon\inst{3} 
\and\\ D.A.~Weintraub\inst{4} 
\and K.\ Hamaguchi\inst{5,6} 
\and A.~Frank\inst{7}}
\institute{Laboratoire d'Astrophysique de Grenoble, Universit{\'e} Joseph-Fourier, Grenoble, F-38041, France
\and Rochester Institute of Technology, Rochester, New York 14623-5604, USA
\and Institute for Astronomy, Honolulu, Hawaii 96822, USA
\and Vanderbilt University, Nashville, Tennessee 37235, USA
\and NASA/Goddard Space Flight Center, Greenbelt, Maryland 20771, USA
\and National Research Council, 500 Fifth Street, NW, Washington, D.C. 20001, USA
\and University of Rochester, Rochester, New York 14627-0171, USA}

  \offprints{Nicolas.Grosso@obs.ujf-grenoble.fr\,.}
   \date{Received ; accepted }
\abstract{We report a $\sim$38\,ks X-ray observation of
   McNeil's Nebula obtained with \xmm~on 2004 April~4. \vor,
   the young star in outburst illuminating McNeil's Nebula,
   is detected with \xmm~and appears variable in X-rays. We
   investigate the hardness ratio variability and time
   variations of the event energy distribution with quantile
   analysis, and show that the large increase of the count
   rate from \vor{} observed during the second half of the
   observation is not associated with any large plasma
   temperature variations as for typical X-ray flares from
   young low-mass stars. X-ray spectral fitting shows that
   the bulk ($\sim75\%$) of the intrinsic X-ray emission in
   the 0.5--8\,keV energy band comes from a soft plasma
   component, with $kT_{\rm soft}=0.9$\,keV (0.7--1.1\,keV,
   at the 90\% confidence limit), reminiscent of the X-ray
   spectrum of the classical T~Tauri star TW Hya, for which
   X-ray emission is believed to be generated by an
   accretion shock onto the photosphere of a low-mass
   star. The hard plasma component, with $kT_{\rm
   hard}=4.2$\,keV (3.0--6.5\,keV), contributes $\sim25\%$
   of the total X-ray emission, and can be understood only
   in the framework of plasma heating sustained by magnetic
   reconnection events. We find a hydrogen column density of
   $N_{\rm H}=4.1\times10^{22}$\,cm$^{-2}$
   (3.5--$4.7\times10^{22}$\,cm$^{-2}$), which points out a
   significant excess of hydrogen column density compared to
   the value derived from optical/IR observations,
   consistent with the picture of the rise of a wind/jet
   unveiled from ground optical spectroscopy. The X-ray flux
   observed with \xmm~ranges from roughly the flux observed
   by \cxo~on 2004 March~22 (i.e.\ $\sim$10 times greater
   than the pre-outburst X-ray flux) to a value two times
   greater than that caught by \cxo~on 2004 March~7 (i.e.\
   $\sim$200 times greater than the pre-outburst X-ray
   flux). The X-ray variability of \vor~in outburst is
   clearly enhanced. We have investigated the possibility
   that \vor{} displays a periodic variation in X-ray
   brightness as suggested by the combined \cxo+\xmm{} data
   set. Assuming that the X-ray flux density is periodic,
   the folding of the two \cxo~observed X-ray flux densities
   with the \xmm~ones leads to three periodic X-ray light curve
   solutions. Our best period candidate is 0.72\,day, which
   corresponds to the time scale of the Keplerian rotation
   at a distance of 1 and 1.4 stellar radius for a one solar
   mass star aged of 0.5 and 1\,Myrs, respectively. We
   propose that the emission measure, i.e.\ the observed
   X-ray flux, is modulated by the Keplerian rotation of the
   inner part of the \vor~accretion disk. 
               }

\maketitle

\section{Introduction}

At the end of January 2004, a new bright fan-shaped nebula
was discovered serendipitously (\cite{mcneil04}) between
M\,78/NGC\,2068 and the region of HH\,24-26. These two well
known star-forming regions of the L1630 dark cloud are
located in the northern part of the Orion B giant molecular
cloud, at a distance of $\sim$400\,pc (\cite{anthony82}). At
the apex of McNeil's Nebula lies the young stellar object
(YSO) \object{IRAS\,05436-0007}=\object{2MASSJ\,05461313-0006048}
(\cite{clark91}), associated with a faint $I$-band source
(\cite{eisloffel97}) detected in the (sub)millimeter
(1.3\,mm source \object{LMZ12} of \cite{lis99}, and
850\,$\mu$m source \object{OriBsmm\,55} of
\cite{mitchell01}). This YSO is now displaying a dramatic
optical/IR outburst, which is the origin of the rise of this
reflection nebula. \cite*{briceno04} have constrained the
start of this outburst to the beginning of November 2003,
and have obtained a light curve in the $I$-band of
\object{\vor} (designation of the variable star
illuminating McNeil's Nebula; \cite{samus04}) showing a
$\sim$5\,mag brightening in about 4 months. They show that
the timescale for the nebula to develop is consistent with
the light-travel time, indicating that we are observing
light from the central source scattered by material in the
cometary nebula.  

The nature of the \vor~outburst and its connection with
other pre-main sequence (PMS) eruptive objects, namely EXors
(stellar prototype: EX\,Lupi; \cite{mcLaughlin46};
\cite{herbig01}), and FUors (stellar prototype: FU\,Ori;
\cite{herbig66}; \cite{herbig77}), is still debated at the
moment. \cite*{reipurth04} and \cite*{mcgehee04}, have noted
a resemblance to EXors; whereas \cite*{briceno04},
\cite*{abraham04}, \cite*{walter04}, and \cite*{kun04} have
proposed a FUor event. \cite*{vacca04} have even reported
that the near-IR spectrum of \vor~does not appear similar to
any known FUor or EXor object. \cite*{aspin04}
reported an $R$-band observation of \vor~ with the Gemini 
telescope on Mauna Kea. These latest data, obtained on 2004
August~29 (shortly after the object emerged from conjunction
with the Sun), indicate that
\vor~continues to be in a state of elevated optical/IR
emission, and that McNeil's nebula remains bright, and
therefore that the duration of the outburst so far exceeds
about nine months. Indeed, continued optical/near-IR
monitoring are necessary to determine the duration of the
outburst, because outbursts last only several months in
EXors compared to several decades in FUors. Both types of
outburst are thought to be driven by a sudden increase of
accretion through a circumstellar disk (e.g.,
\cite{hartmann96} and references therein), but the
distinction between FUors and EXors is still entirely
empirical. The interpretation that the outburst event has
its origin in accretion processes is supported by
high-resolution IR spectra of \vor{} showing CO emission
lines, likely originating from $\sim 2500$\,K gas in an
inner accretion disk region where substantial clearing of
dust has occurred (\cite{rettig05}). 

The pre-outburst spectral energy distribution (SED) of
\vor~from IR to millimeter shows a flat-spectrum source
(\cite{abraham04}). This kind of flat SED is usually
interpreted in terms of a circumstellar envelope
(\cite{kenyon91}). \cite*{andrews04} propose that \vor~is a
transition object between a protostar with circumstellar
disk plus remnant circumstellar envelope (Class~I
protostar), and a classical T~Tauri star with
circumstellar disk (Class~II sources). {\sl Spitzer}
observations of \vor{} in early March 2004
(\cite{muzerolle05}) show a factor of 15--20 increase in
brightness across the spectrum from the optical to
70\,$\mu$m, leading to a bolometric luminosity of
44\,L$_\odot$, i.e.\ $\sim$15 times higher than the
pre-outburst level. 

X-ray imaging spectroscopy observations with \cxo~have 
revealed a factor $\sim$50 increase in the X-ray count 
rate from \vor~during its outburst compared to the pre-outburst 
state (\cite{kastner04}). The coincidence of a surge in X-ray 
brightness with the optical/IR outburst demonstrates that strongly 
enhanced high energy emission from \vor~occurs as a consequence 
of high accretion rates. The burst of X-rays was most probably 
generated via star-disk magnetic reconnection events that occurred 
in conjunction with such mass infall. This process may also launch 
new, collimated outflows or jets (\cite{goodson97}). Indeed, before 
its recent eruption, \vor~had been identified as the exciting source 
of a chain of extended emission nebulosity that appears to terminate 
at HH\,23, a shock-excited Herbig-Haro object, located $\sim3^\prime$ 
North from \vor~(\cite{eisloffel97}; \cite{lis99}; \cite{reipurth04}). 
The presence of these structures suggests that the present 
optical/IR/X-ray outburst of \vor~may be merely the latest of a series 
of such events. Another outburst may have occurred about 37\,yrs 
before the present event, based on a photograph obtained in Oct.\ 1966 
which shows a similar cometary nebula (\cite{mallas78}). No clear 
evidence for the presence of a molecular outflow has been found in 
the submillimeter CO spectral line maps (\cite{lis99}; \cite{andrews04}). 
However the H$\alpha$ line, which has been detected in strong emission, 
displays a pronounced P~Cygni profile, with an absorption trough reaching 
velocities up to 600\,km\,s$^{-1}$ (\cite{reipurth04}), implying 
significant mass loss in a powerful wind.

We report here an X-ray observation of \vor{} in outburst 
obtained with \xmm, showing enhanced X-ray variability from V1647\,Ori. 
We present in \S\ref{obs} the \xmm~observation and data reduction, 
we study the variability in \S\ref{variability}, and the X-ray 
spectrum in \S\ref{spectrum}. We compare the X-ray flux observed 
with \cxo~and \xmm~in \S\ref{comparison}, and we discuss in 
\S\ref{discussion} the origin of the X-ray flux variations observed from \vor.

\section{XMM-Newton observations and data reduction}
\label{obs}

We observed \vor~with \xmm~(\cite{jansen01}) on 2004 April
4. The pointing nominal J2000 coordinates were 
$\alpha_{\rm J2000}$=05$^{\rm h}$46$^{\rm m}$13$\fs$1, 
$\delta_{\rm J2000}$=-00$^\circ$06$^\prime$04$\farcs$6. 
We used the full frame science mode of the EPIC cameras with medium
optical blocking filter. The total exposure times were
$\sim$37.0\,ks and $\sim$38.7\,ks for the EPIC/PN
(\cite{strueder01}) and the two EPIC/MOS spectro-imagers
(\cite{turner01}), respectively. We also acquired optical
and UV data with the Optical Monitor (OM) with the following
filter sequence: $V$, $U$, $B$, and $UVW1$; however the OM
data are severely contaminated by stray light
(\cite{mason01}) from the extended reflection nebula M\,78, 
which is located $\sim$10$\arcmin$--$\sim$15$\arcmin$ North-East
of \vor, preventing us from using these data at the position
of \vor. We will focus here only on the X-ray detection of
\vor. 

The data were reduced using the \xmm{} Science Analysing
System (SAS, version 6.0.0)\footnote{For a description of
the standard procedures see, e.g., the \xmm{} ABC Guide,
Snowden et al.\ (2004), available at 
{\tt http://heasarc.gsfc.nasa.gov/docs/xmm/abc/\,.}}. The 
event lists were produced using tasks {\tt epchain} and 
{\tt emchain}. The observation is mainly affected by a
background proton flare having a duration of one hour,
centered at 1\,h U.T.\ on 2004 April 4, and peaking on
PN\footnote{Background value obtained from the detector
background light curve provided for PN by {\tt epchain}.}
to 50--150\,count\,ks$^{-1}$\,arcmin$^{-2}$. Removal of
time intervals with flaring background above 10 and
5\,count\,ks$^{-1}$\,arcmin$^{-2}$ for PN and MOS,
respectively, leads to effective exposure times of $\sim$33\,ks
and $\sim$38\,ks, respectively.  X-ray images were then
produced by selecting good events within the fields-of-view
of the detectors\footnote{We selected single, double,
triple, and quadruple pixel events ({\tt PATTERN} in the 0
to 12 range) for both PN and MOS, plus the {\tt FLAG} list
{\tt \#XMMEA\_EM} for MOS.}.  

\begin{figure}[!t]
\centering
\includegraphics[width=\columnwidth]{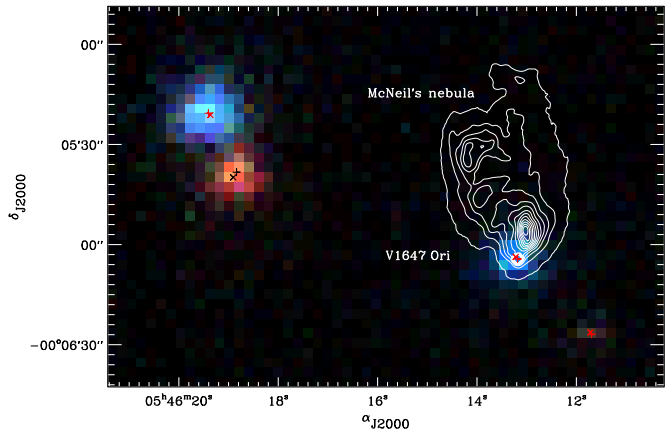}
\caption{\xmm/EPIC~image of the region surrounding McNeil's
  Nebula. The three images obtained with PN, MOS1, and MOS2
  EPIC cameras on 2004 April 4 were added after correction
  from vignetting. The size of the image pixel is
  3$\arcsec$. This image has red, green, and blue colour
  coding for X-ray photons in the 0.5--1.5, 1.5--2.4, and
  2.4--8.0\,keV energy bands, respectively (see algorithm of
  Lupton et al.\ 2004). For comparison purposes we overlay a
  $R$-band contour map of the nebula obtained with {\it
    VLT/FORS2} on 2004 February 18. X-ray sources and 2MASS
  sources are marked with `$\times$' and `+' respectively. A
  bright embedded X-ray source is spatially coincident with
  \vor~at the apex of McNeil's nebula.}  
\label{trichro}
\end{figure}

Vignetting corrections were made from exposure maps
computed on the corresponding energy
bands. Figure~\ref{trichro} shows the tricolour X-ray image
of the region surrounding McNeil's Nebula (see for
comparison Fig.~2 of \cite{kastner04}). The X-ray source
\object{XMM\,J054613.1-000604} found by the pipeline is
located at $\alpha_{\rm J2000}$=05$^{\rm h}$46$^{\rm
  m}$13$\fs$2, $\delta_{\rm
  J2000}$=-00$^\circ$06$^\prime$04$\farcs$1 with a $1\sigma$
error box of 0.2$\arcsec$, and displays a count rate of
$0.036\pm0.001$\,count\,s$^{-1}$ in PN and
$0.012\pm0.001$\,count\,s$^{-1}$ in each MOS detector within the
energy band 0.2--12\,keV. This X-ray source is located at
$0.9\arcsec$ from the 2MASS position of \vor. Computing the
root-mean-square dispersion of distances between the
positions of X-ray sources found by the pipeline and their
2MASS counterpart positions, we found the same offset value,
therefore we will consider this \xmm~source to be the same
X-ray counterpart of \vor~already detected by
\cxo~(\cite{kastner04}). The other X-ray sources visible in
Fig.~\ref{trichro} are (from left to right):
\object{XMM\,J054619.4-000520},
\object{XMM\,J054618.9-000539}, and
\object{XMM\,J054611.6-000627}, which are the X-ray counterparts to
\object{SSV\,64}=\object{LkH$\alpha$301},
\object{GSC2\,S020001258}, and
\object{2MASS\,J05461162-0006279}, respectively (see \cite{kastner04}, 
and \cite{simon04}).  

We selected in each camera X-ray events of the \vor~X-ray
source for variability and spectral analysis. For PN, we
selected only single and double pixel events (i.e.\ 
{\tt PATTERN} in the 0 to 4 range) with {\tt FLAG} value equal
to zero. Source plus background events were extracted within a 
circular region using the SAS task {\tt eregionanalyse}. 
The extraction position and radius ($r=23\arcsec$) 
were optimized by this task to maximize the signal-to-noise
ratio, taking for the estimation of the background an
annular region  ($r=23\arcsec$--$44\arcsec$) centered on
this source and located on the same CCD, where areas
illuminated by other X-ray sources were excluded. Background
events of this annular region were also extracted. 

\section{Variability and quantile analysis}
\label{variability}

We limited our variability analysis to the PN observing time
interval, where all three EPIC cameras were working
together, which allowed us to build directly an EPIC
(PN+MOS1+MOS2) light curve. Using the SAS task {\tt 
evselect}, we first built two light curves with 1\,s
time bins from the source+background and
the background event lists (where we kept flaring background
time intervals). These light curves begin 
exactly at the start of the first Good Time
Interval (GTI) of the PN aimpoint CCD \#4 (where the source
is detected). We rebinned the extracted light curve to a
longer time interval with the ftools task {\tt lcurve} to
increase the signal. Then, we subtracted from the
source+background light curve the background light curve
scaled to the same source extraction area, using the ftools
task {\tt lcmath}. We also corrected this source light curve
for the observing time lost, for example, due to the triggering
of counting mode during high flaring background periods, when
the count rate exceeded the detector telemetry limit. From the GTI
extension, we computed with an IDL routine for each time bin
the ratio between the time bin length and the observing time
lost in this time bin, and we multiplied count rate and
error by this linear correction factor. Finally, light
curves of the three detectors were summed to produce the EPIC
light curve. 

\begin{figure}[!t]
\centering
\includegraphics[width=\columnwidth]{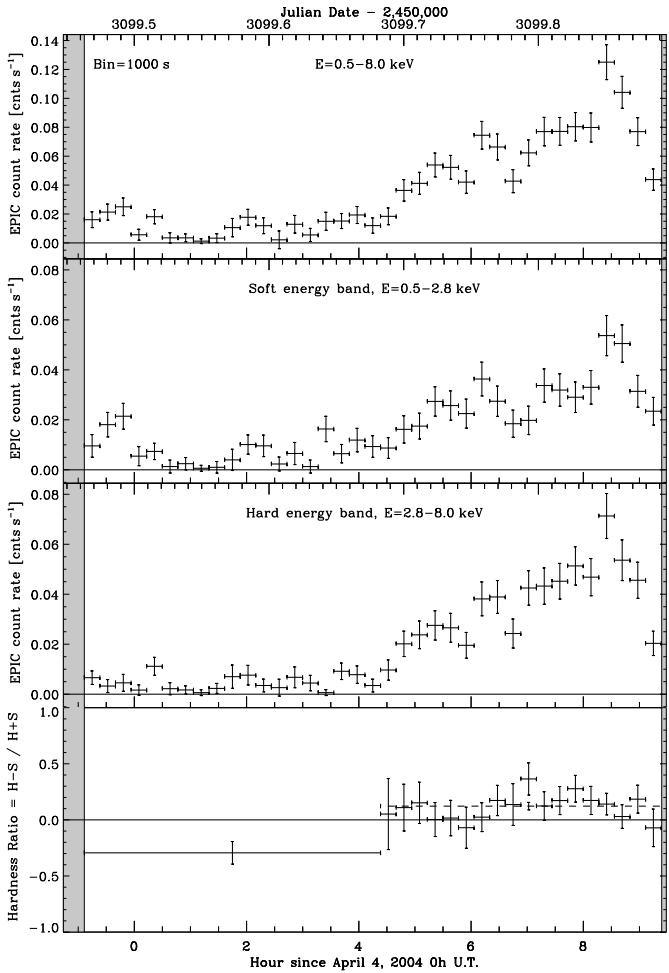}
\caption{\xmm/EPIC background subtracted X-ray light curves
  of V1647\,Ori. The upper panel shows the EPIC
  (PN+MOS1+MOS2) X-ray light curve of \vor~with one sigma
  error bars in the energy band from 0.5 to 8.0\,keV. The
  grey stripes indicate the small observational time
  intervals before the start and the end of the PN exposure
  where only the two MOS detectors were observing. The middle
  panels show the soft ($S$=0.5--2.8\,keV) and hard
  ($H$=2.8--8\,keV) band X-ray light curves. The lower panel
  shows the variation of the corresponding hardness
  ratio. The dashed line indicates the hardness ratio for
  the second half of the observation. 
}
\label{lc}
\end{figure}

The upper panel of Figure~\ref{lc} shows the EPIC X-ray
light curve of \vor~in the energy band from 0.5 to 8.0\,keV
binned to 1000\,s.  The source is variable, with a low level 
in count rate during the first half of the observation,
varying between $\sim$0.025 and $\sim$0.005\,count\,s$^{-1}$
(note that the decrease of count rate around 1\,h U.T.\
cannot be a consequence of the high flaring background
because we have corrected for the losses of observing time
due to triggering of counting mode). This low level was then
followed by a much higher level of count rate during the
second half of the observation. This latter phase consisted of a slow
($\sim$4\,h) rise with an abrupt jump to a peak at
$\sim$0.13\,count\,s$^{-1}$, and then a faster ($\sim$1\,h)
decay from the peak of $\sim$0.13\,count\,s$^{-1}$ back down
to the general level of enhanced emission seen from 5\,hr
onwards. A similar time behaviour is visible both in the
soft energy band ($S$=0.5--2.8\,keV) and in the hard energy
band ($H$=2.8--8\,keV). To quantify possible softening or
hardening of the source events with time, we use the
standard hardness ratio method by computing $HR=(H-S)/(H+S)$
(where the hardness ratio errors are computed using Gaussian error
propagation). The hardness ratios for the first and the
second halves of the observation are $HR=-0.29 \pm 0.10$ and
$0.12\pm0.03$, respectively (see the lower panel of
Figure~\ref{lc}).  

\begin{figure}[!t]
\centering
\includegraphics[width=\columnwidth]{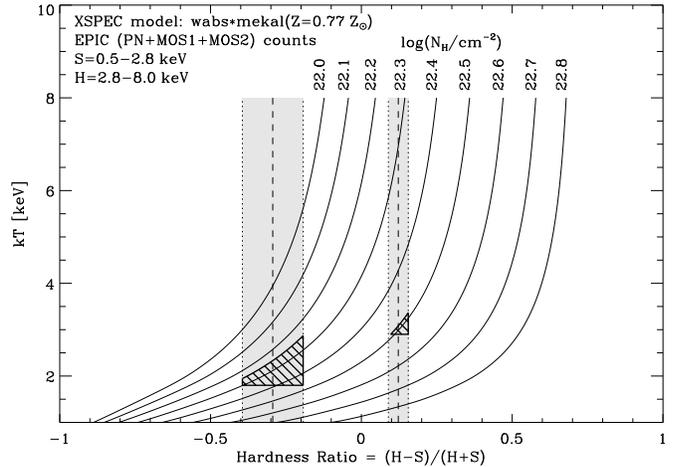}
\caption{Comparison between the observed hardness ratios and
  absorbed one temperature plasma models. The left and right
  hand vertical grey stripes indicate the hardness ratio
  values of the first and the second halves of the
  observation, respectively, measured with EPIC (see lower
  panel of Fig.~\ref{lc}). The curves give plasma
  temperature versus hardness ratio for a fixed value of the
  hydrogen column density. The dashed areas show the
  constraints on the model parameters when combining the
  results from the quantile analysis of PN data (see lower panel of
  Fig.~\ref{quantiles}) with hardness ratios
  measured with EPIC.} 
\label{hr_modeling}
\end{figure}

A least-square fit assuming constant hardness ratio leads to 
$\chi^2 = 15.4$ (with one degree of freedom), with a 
probability\footnote{Here, $Q$ is the probability that the 
best-fit model matches the data, given the value of $\chi^2$. 
Values of $Q < 0.05$ indicate that the best-fit model is a poor 
fit, while $Q \sim 1$ indicates the fit is overconstrained. 
Optimal values of $Q$ are $\sim0.5$.} of $Q\sim10^{-4}$. This 
fit therefore excludes, at the 99.99\% confidence level, the 
possibility that the hardness ratio remained constant during the 
\xmm{} observation. We conclude that the X-ray emission from \vor{} 
hardened during the second half of the observation.

However during the second half of the observation, there is
no correlation between the observed variations of the full
energy band light curve and the hardness ratio. The hardness
ratio values computed within time bins of 1000\,s (see
the lower panel of Figure~\ref{lc}) are consistent with a
constant level (with a probability of 89\%). In particular,
the rise and the decay of the light curve do not correspond
to any hardening and softening of the spectrum, whereas such
a correspondence is usually observed in typical X-ray flares
from young low-mass stars that are associated with the heating and the
cooling phases of the emitting plasma (e.g.,
\cite{tsuboi98}, \cite{imanishi03}); in such cases, the absorbing column
density is observed to remain constant.

\begin{figure}[!t]
\centering
\includegraphics[width=0.9\columnwidth]{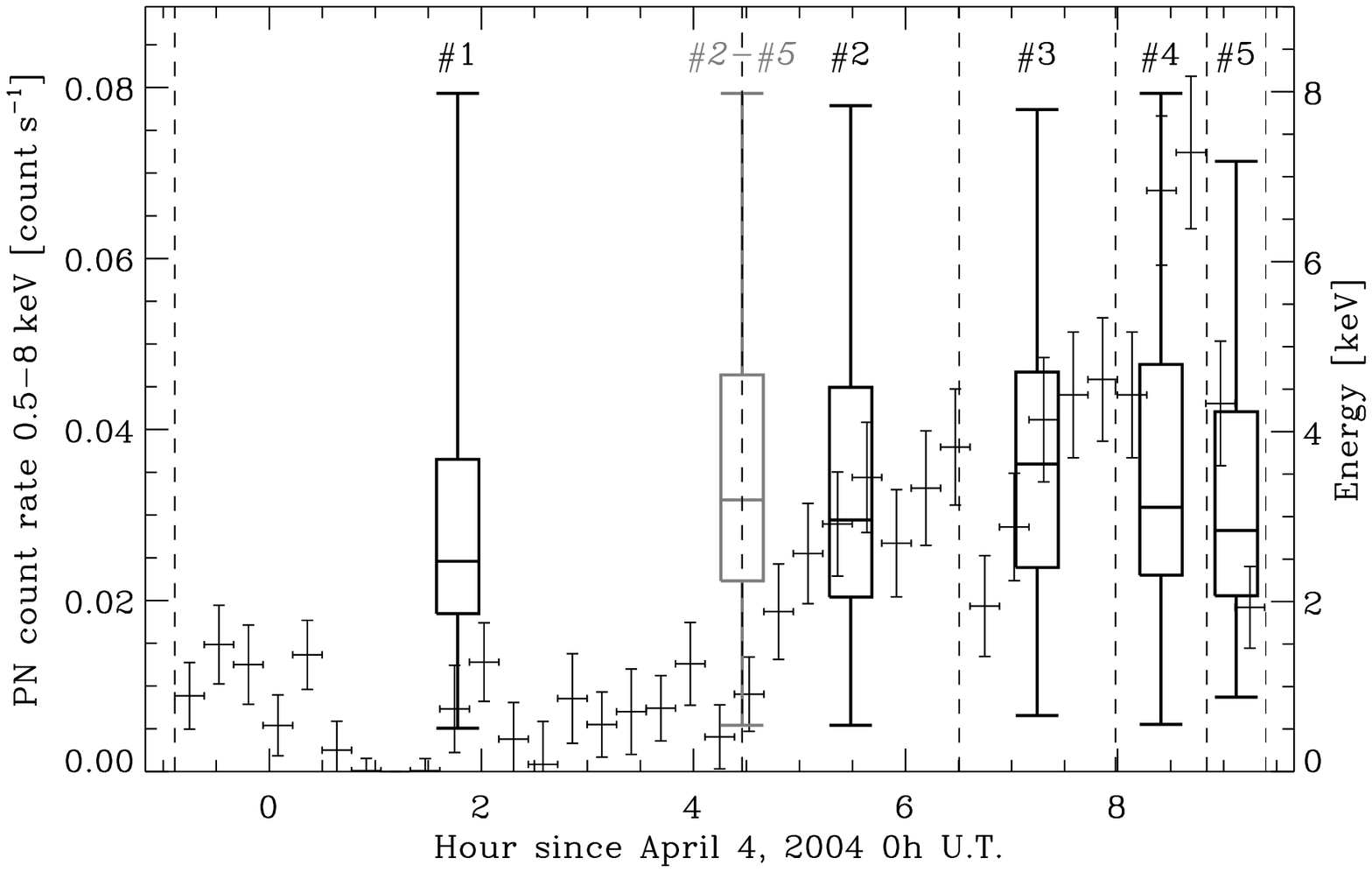}
\includegraphics[width=0.9\columnwidth]{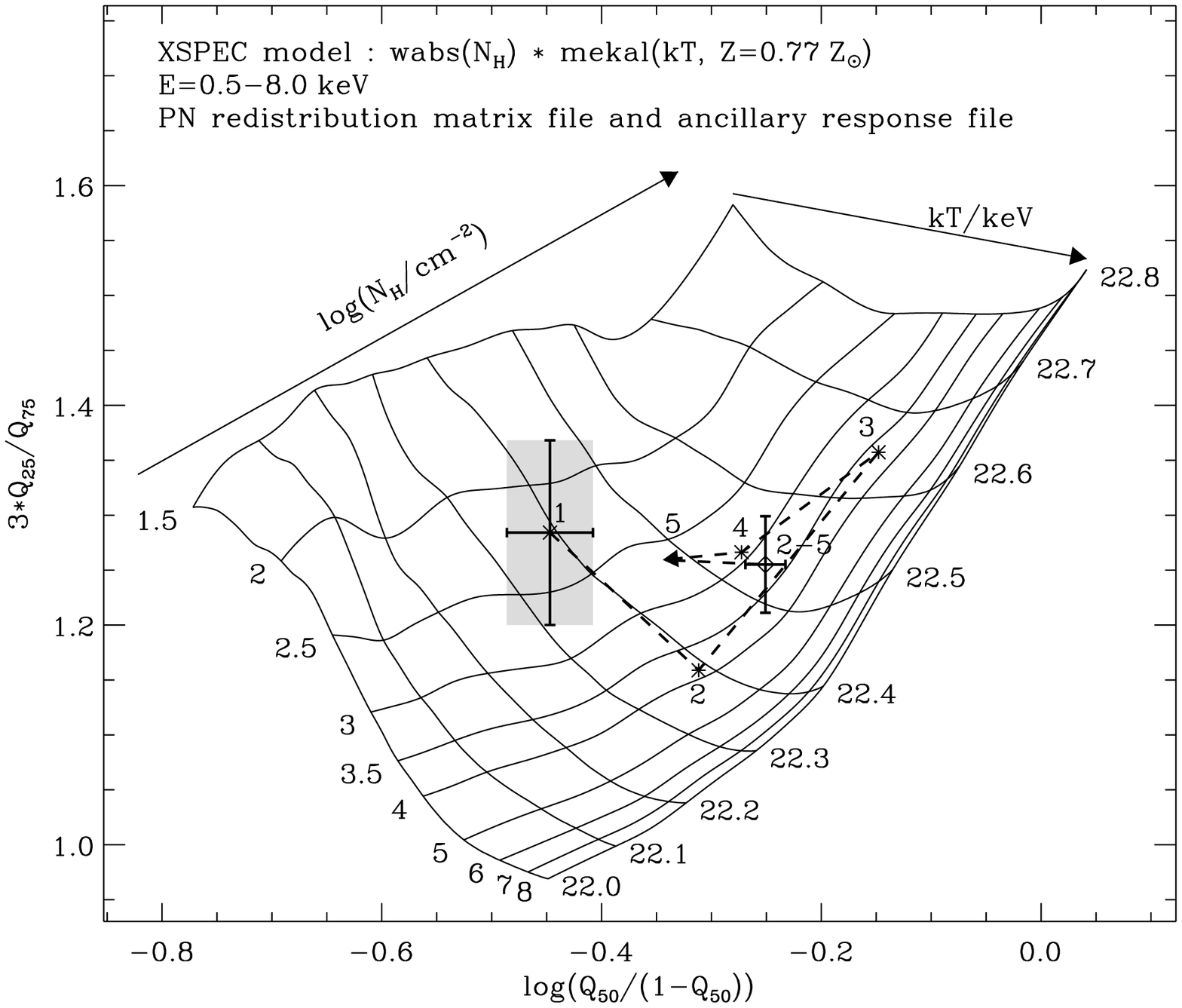}
\caption{Quantile analysis of the energy of the X-ray photons 
of \vor~collected by EPIC/PN. The upper panel shows the EPIC/PN 
light curve of \vor~in the 0.5--8.0\,keV energy band. Vertical 
dashed lines define variable-length time bins numbered from \#1 to \#5 
where PN collected 200 net counts. For each time bin, the box plot 
indicates the energy distribution of the source (the vertical line and 
the box indicate the energy range and the quartiles/median, respectively) 
corrected from background (except minimum and maximum values) using the 
Hong et al.\ (2004) method. The grey box plot corresponding to bin \#2--\#5 
indicates the energy distribution of the entire second half of the observation. 
The lower panel shows the X-ray colour-colour diagram based on median and the 
ratio of two quartiles (Hong et al.\ 2004). The numbered asterisks mark the 
X-ray colours corresponding to the time bins defined in the upper panel. 
The diamond marks the X-ray colours of the second half of the observation. 
Typical one sigma error bars are shown for comparison.}
\label{quantiles}
\end{figure}

Figure~\ref{hr_modeling} shows a comparison of the
hardness ratio values measured with EPIC for the first and
the second halves of the observation with absorbed one
temperature plasma models computed with XSPEC (version
11.3.0)\footnote{XSPEC {\tt wabs}$\times${\tt mekal} model,
  with global metallicity abundance $Z=0.77$\,Z$_\odot$
  (obtained from spectral fitting; \S\ref{spectrum}), using
  the PN, MOS1, and MOS2 redistribution matrix files and
  ancillary response files appropriate for \vor~(on-axis
  source).}. We cannot disentangle possible changes in
absorbing column density from changes in plasma temperature
with only one hardness ratio. The usual improvement would be
to introduce three consecutive energy bands, which would
allow the computation of two hardness ratios: the first
built on soft and medium energy bands, sensitive to the
absorbing column density; the second built on medium and
hard energy bands, sensitive to the plasma
temperature. However the total number of counts is too limited
to permit us to obtain two hardness ratios with small enough
error bars to allow a temporal study. To improve our
analysis, we investigate changes in the energy of the source
X-ray photons collected by PN from quantile analysis, as
proposed by \cite*{hong04}. 

To avoid signal-to-noise variations, we define
variable-length time bins in which the total number of net
counts collected by PN is constant, and equal to 200 net
counts per time bin. This leads to 5 time bins(where bin \#1
corresponds to the first half of the observation) and
thereby allows a variability study. Following
\cite*{hong04}, we determine for each time bin median
($E_{50\%}$) and quartile ($E_{25\%}$, $E_{75\%}$) energies
of the source+background X-ray photons collected by PN, 
corrected for the background X-ray photons (see upper
panel of Fig.~\ref{quantiles}); and associated quantiles,
$Q_{\rm x\%}=(E_{\rm x\%}-E_{\rm low})/(E_{\rm up}-E_{\rm low})$, 
with $E_{\rm low}=0.5$\,keV and $E_{\rm up}=8$\,keV. 
Then, we compute the two X-ray colours
proposed by \cite*{hong04}: $x \equiv \log (Q_{50\%}/(1 -
Q_{50\%}))$, and $y \equiv 3\times Q_{25\%}/Q_{75\%}$. These
definitions of X-ray colours lead to a unique correspondance
between the X-ray colour pair ($x$,$y$) and plasma parameters
($N_{\rm H}$,$T$), where $N_{\rm H}$ is the absorbing
hydrogen column density, and $T$ the plasma temperature. The
lower panel of Fig.~\ref{quantiles} shows the reference
mapping of the $x$,$y$ space-parameter that we obtained by
simulating with XSPEC absorbed one-temperature plasma
spectra from an on-axis source observed with PN. This grid
allows a direct estimate of $\log N_{\rm H}$ and $kT$ from
$x$ and $y$. Based on 10,000 trials of synthetic spectra
computed from a model with parameters $\log N_{\rm H}=22.5$,
$kT=3.0$\,keV, $Z=0.77$\,Z$_\odot$ (see \S\ref{spectrum} and
model \#1 in Table~\ref{fit_parameters}), and 200 counts, we
found that the errors on $x$ and $y$ colours are distributed
normally. In the lower panel of Fig.~\ref{quantiles}, we
plot one sigma error bars on $x$ and $y$ for a single (200
net count) time bin.  

For bin \#1, taking into account the one sigma uncertainties
on the X-ray colours, we find $\log N_{\rm H} \sim 22.4$
(22.25--22.5), and $kT \sim 2.2$\,keV (1.8--2.8\,keV). Over
the course of the entire observation, the $y$ colour varied
by less than the one sigma uncertainty for a single time
bin. From bin \#1 to bins \#3 and \#4, there is a variation
of the $x$ colour greater than the one sigma error
bar. However, due to the large error on the $y$ colour, this
variation in $x$ colour implies no significant variation of
the hydrogen column density and/or the plasma
temperature. During the second half of the
observation, we find
$\log N_{\rm H} \sim 22.53$ (22.49--22.57) and $kT \sim 3.2$\,keV
(2.9--3.7\,keV); because $\sim80$\% of the photons were
detected in this latter half of the observation, the uncertainties
on the quantile values obtained for the second half of the
observation are smaller by a factor $\sim2$ than the 
uncertainties in the first half of the observation.
Therefore, there appears to be a hint of an
increase in plasma temperature during the second half of the
observation.

In Figure~\ref{hr_modeling}, we combine these constraints on
the hydrogen column density and the plasma temperature of
the first and the second halves of the observation with the
hardness ratio values obtained previously. We conclude that
the hardening of the spectrum can be explained by a rather
small increase of the plasma temperature from $\sim$2\,keV
to $\sim$3\,keV, combined with a small increase of the
absorbing column density from $\log N_{\rm H} \sim 22.3$ to
$\log N_{\rm H} \sim 22.5$. 

It appears overall that the large increase of the count rate
is not associated with any large variations of the
temperature. This lack of strong correlation between X-ray
flux and temperature suggests that we are perhaps also observing
variations in the emission measure of the X-ray-emitting
plasma. The variability 
observed with \xmm~and that measured previously with
\cxo~on 2004 March 7 and 22 will be compared further in
\S\ref{comparison} to investigate this possibility.

\section{Spectral analysis}
\label{spectrum}

\begin{figure}[!t]
\centering
\includegraphics[width=0.78\columnwidth,angle=-90]{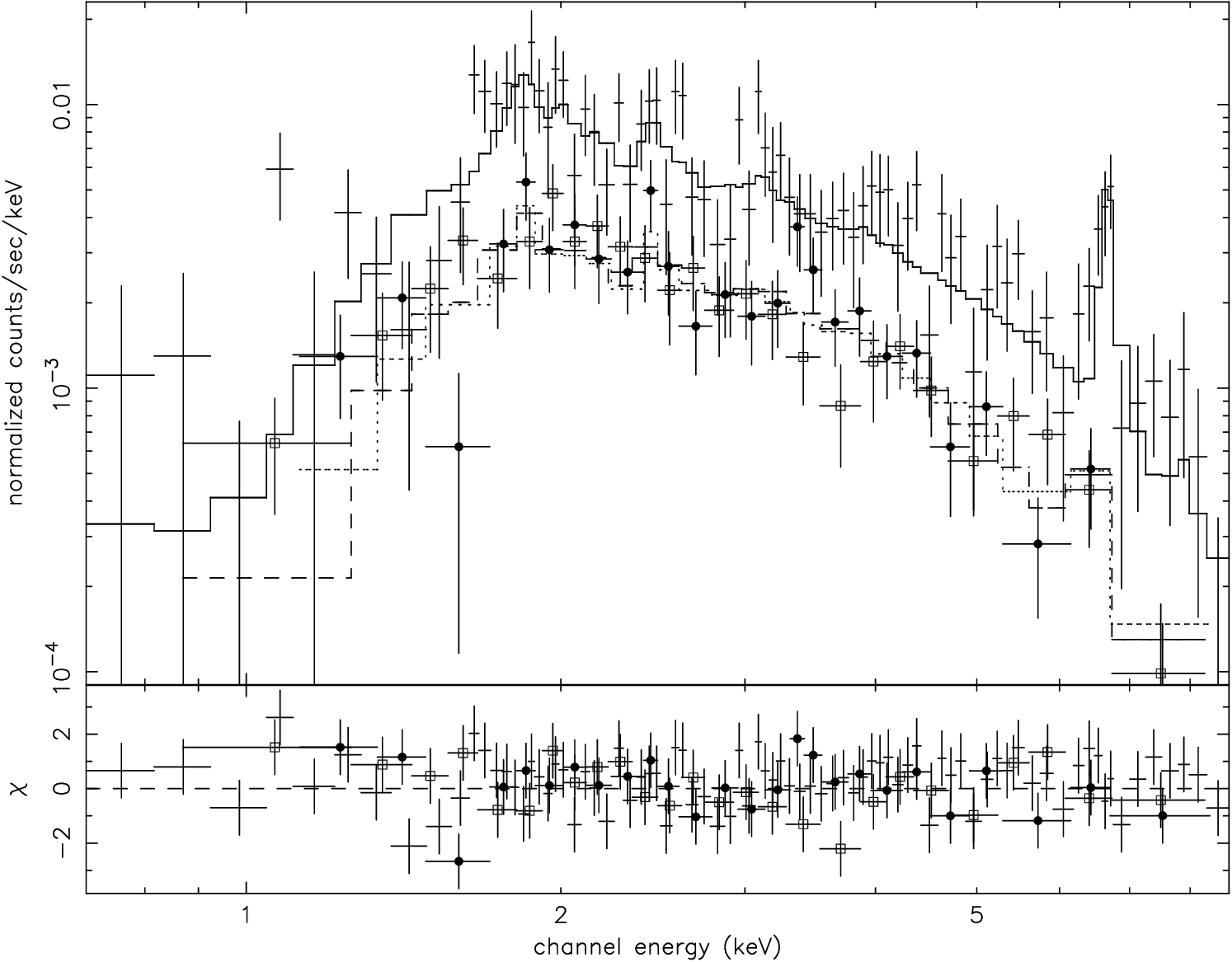}
\caption{\xmm~X-ray spectra and our best fit models of V1647\,Ori. 
The upper panel shows the data binned to 15 counts in each spectral 
bin, and our best fit models with an absorbed two-temperature 
optically thin plasma (see model \#2 in Table~\ref{fit_parameters}). 
Continuous line, dashed line plus squares, and dotted line plus disks, 
stand for PN, MOS1, and MOS2, respectively. The lower panel shows 
the residuals of the fit, in units of the uncertainties in the 
individual data points. The Fe\,{\small XXV} emission line at 
6.7\,keV is clearly visible.
}
\label{spectra}
\centering
\begin{tabular}{@{}cc@{}}
\includegraphics[width=0.61\columnwidth,angle=-90]{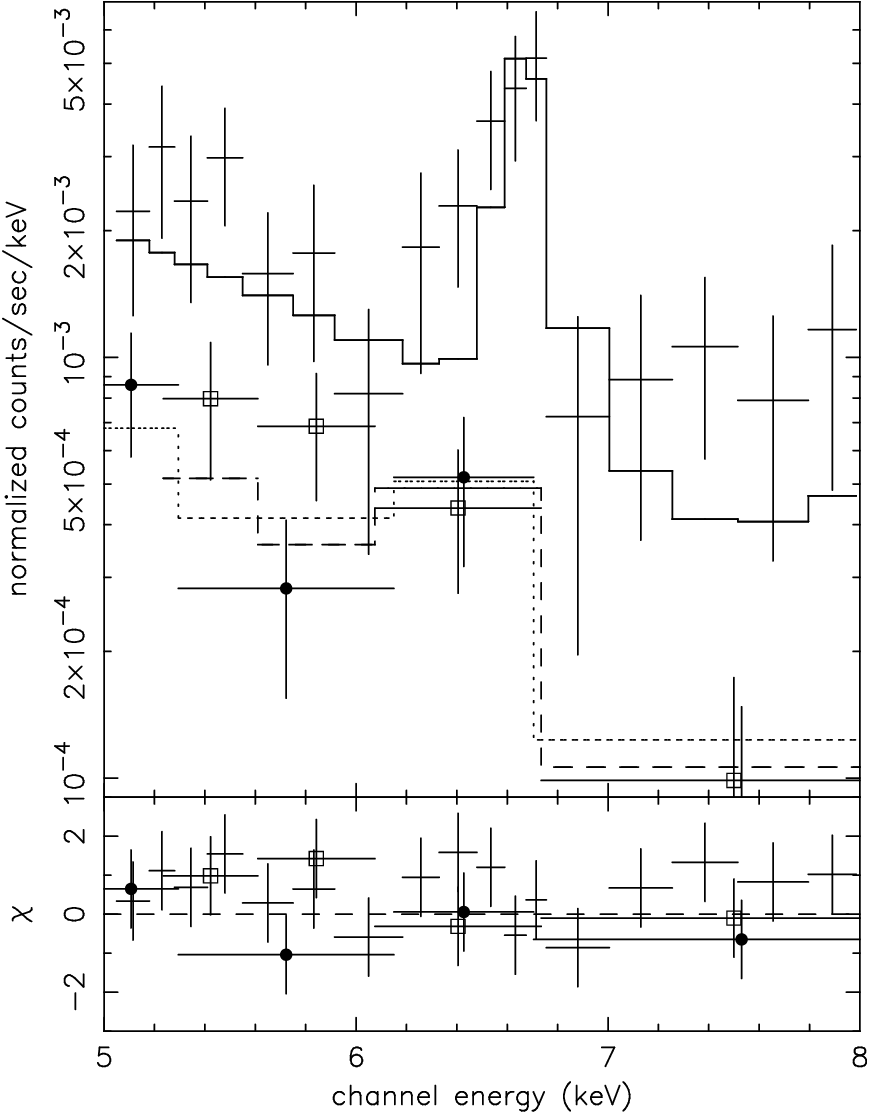} &
\includegraphics[width=0.61\columnwidth,angle=-90]{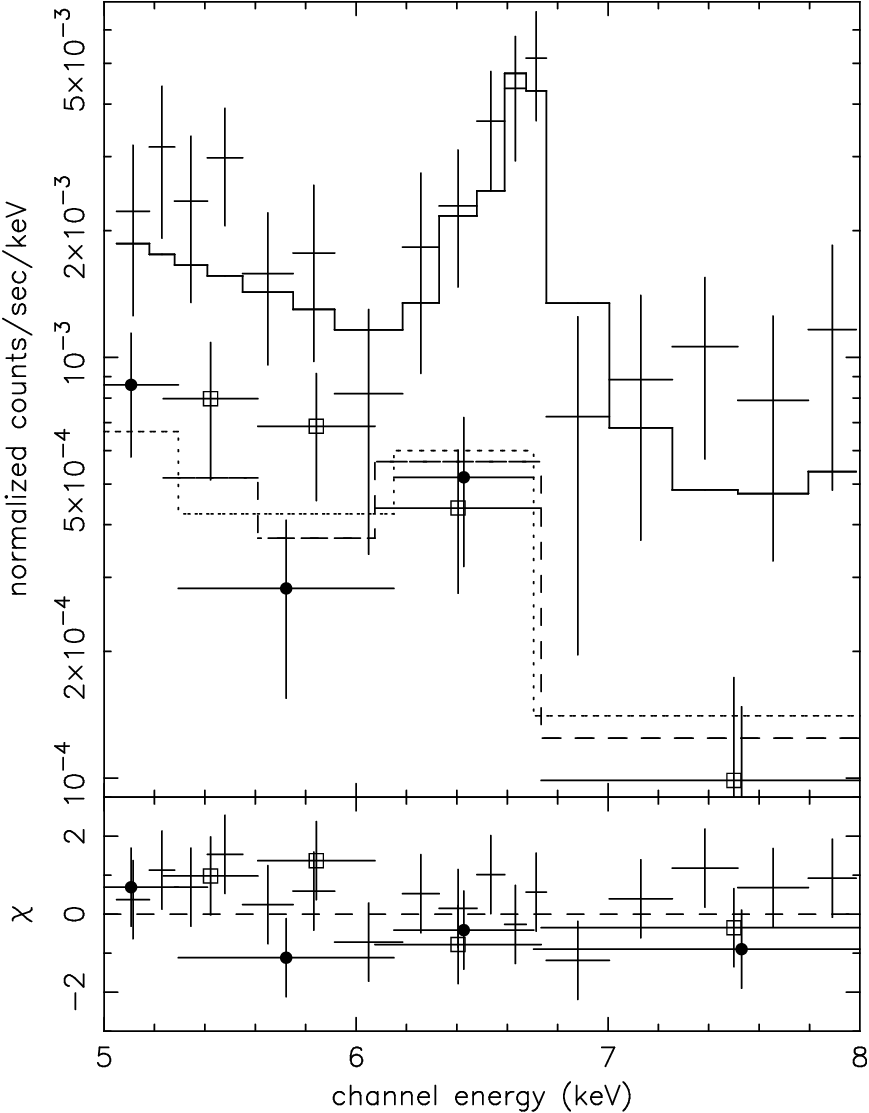}
\end{tabular}
\caption{Fit around the 6.7\,keV iron line region. The left hand 
panel shows an enlargement of the absorbed two-temperature model 
(see model~2 in Table~\ref{fit_parameters}) around the 6.7\,keV 
iron line, using a linear energy scale. The right hand panel shows 
the same energy area adding a Gaussian line fitting the 6.4\,keV 
neutral iron line (see model 3 in Table~\ref{fit_parameters}). 
Symbols are identical to the ones of Fig.~\ref{spectra}.
}
\label{line}
\end{figure}

\begin{table*}[!t]
\caption{Best fit parameters of EPIC spectrum models with their 
errors at the 90\% confidence levels ($\Delta\chi^2=2.71$; 
corresponding to $\sigma=1.64$ for Gaussian statistics). For PN, 
MOS1, MOS2, the effective exposures after suppression of time 
intervals with high levels of flaring background are 33.2, 37.7, 
and 38.1\,ks, and there are in the source extraction region 1264, 
408, 408 counts, respectively ($\sim20\%$ of these counts come from the background).}
\label{fit_parameters}
\begin{tabular}{@{}ccccccccrr@{}r@{}rc@{}}
\hline
\hline
\noalign{\smallskip}
 & & & \multicolumn{2}{c}{Temperature}  &\multicolumn{2}{c}{Emission Measure$^\mathrm{a}$} & \multicolumn{2}{c}{6.4\,keV Fe Line$^\mathrm{b}$} & & & & \multicolumn{1}{c}{$L_\mathrm{X,intr.}$$^\mathrm{a}$}\\
\vspace{-0.6cm}\\
 & & & \multicolumn{2}{c}{\hrulefill}   & \multicolumn{2}{c}{\hrulefill} & \multicolumn{2}{c}{\hrulefill} & & & &  \multicolumn{1}{c}{\hrulefill} \\
Model & $N_\mathrm{H,22}$& $Z$   & \multicolumn{1}{c}{$T_\mathrm{soft}$} & \multicolumn{1}{c}{$T_\mathrm{hard}$} & \multicolumn{1}{c}{$EM_\mathrm{soft}^\mathrm{54}$} & \multicolumn{1}{c}{$EM_\mathrm{hard}^\mathrm{54}$}  & $N_{\rm ph,-7}$ & $EW$ & \multicolumn{2}{c}{$\chi^2_\nu$ ($\nu$)$^\mathrm{c}$} & \multicolumn{1}{c}{$Q\,^\mathrm{d}$} & 0.5--2.8/2.8--8/0.5--8\\
\# & [cm$^{-2}$]   & [$Z_\odot$] & \multicolumn{2}{c}{[keV]}  & \multicolumn{2}{c}{[cm$^{-3}$]} & [cm$^{-2}$\,s$^{-1}$] & [eV] & & &\multicolumn{1}{c}{[\%]} & \multicolumn{1}{c}{$\log$\,[erg\,s$^{-1}$]}    \\
\noalign{\smallskip}
\hline
\noalign{\smallskip}
1 & 2.9$_{-0.4}^{+0.5}$ & 0.8$_{-0.3}^{+0.5}$ & \dotfill            & 3.0$_{-0.6}^{+0.9}$ & \dotfill            & 0.7$\pm$0.1  & \dotfill & \dotfill       & 1.11 &~(129) & 19 & 30.8 30.5 31.0\\
\noalign{\smallskip}
2 & 4.1$\pm$0.6 & 0.8$_{-0.3}^{+0.4}$ & 0.9$\pm$0.2 & 4.3$_{-1.3}^{+2.2}$ & 1.2$_{-0.6}^{+1.3}$ & 0.4$_{-0.1}^{+0.2}$  & \dotfill & \dotfill       & 0.97 &~(127) & 56 & 31.4 30.5 31.5 \\
\noalign{\smallskip}
3 & 4.1$\pm$0.6 & 0.8$_{-0.3}^{+0.4}$ & 0.9$\pm$0.2 & 4.2$_{-1.3}^{+2.4}$ & 1.3$_{-0.6}^{+1.4}$ & 0.4$_{-0.1}^{+0.2}$  & 4.3$_{-4.3}^{+4.5}$ &  109      & 0.96 &~(126) & 60 & 31.4 30.5 31.5 \\
\noalign{\smallskip}
\hline
\end{tabular}
\begin{list}{}{}
\small
\item[$^{\mathrm{a}}$] $EM$ and $L_\mathrm{X,intr.}$ were 
computed assuming 400\,pc for the distance of V1647\,Ori. 
The X-ray intrinsic luminosity was estimated from the fit 
of the PN data.
\item[$^{\mathrm{b}}$] Energy and width fixed to 6.4\,keV 
and 12\,eV, respectively.
\item[$^{\mathrm{c}}$] $\nu$ is the degree of freedom, 
and $\chi^2_\nu$ is the reduced Chi-square (i.e.\  the 
Chi-square divided by the degree of freedom).
\item[$^{\mathrm{d}}$] The $Q$-value is the probability 
that one would observe the Chi-square value, or a larger 
value, if the assumed model is true, and the best-fit model 
parameters are the true parameter values.
\end{list}
\end{table*}

We created source and background spectra with associated
redistribution matrix files and ancillary response files
using the SAS task {\tt especget} applied on the same source
and background extraction region used for the light curve,
and we binned the spectra to 15 counts per spectral bin using
the ftools task {\tt grppha}. Fig.~\ref{spectra} shows PN,
MOS1, and MOS2 spectra of the whole observation after
suppression of high flaring background periods. The spectrum
is clearly absorbed, with only a few counts detected by PN
below 1\,keV in energy, whereas PN detected  counts up to
9\,keV in energy. A 
prominent line is visible in PN spectra around 6.7\,keV,
corresponding to the Fe\,{\small XXV} triplet emission
line. We performed spectral modeling with XSPEC (version
11.3.0) using a multi-temperature plasma
({\tt mekal}) model (\cite{kaastra96}) with intervening absorption
(as represented by the {\tt wabs} model, which uses
\cite{morrison83} cross-sections). Table~\ref{fit_parameters} gives the best
fit parameters that we found for each statistically
acceptable model (i.e.\ each model leading to a fit with $Q$-value
greater than 5\%), with the corresponding 90\% confidence level
intervals.  

The absorbed one-temperature plasma model ({\tt wabs$\times$mekal}) 
leads to $N_{\rm H}=2.9\times10^{22}$\,cm$^{-2}$ 
(2.5--3.4$\times10^{22}$\,cm$^{-2}$), $kT=3.0$\,keV (2.4--3.8\,keV), 
with $Z=0.8$\,Z$_\odot$ (0.4--1.2\,Z$_\odot$), i.e.\ column density 
and temperature values consistent with the ones previously determined 
from our quantile analysis in \S\ref{variability}. This quantile 
analysis remains the only time-dependent spectral analysis possible 
with this data set due to the small number of total counts.

We obtain a fit with a better probability ($Q$=56\%) when we
add a second temperature component. Therefore, we consider
this model (\#2 in Table~\ref{fit_parameters}) as our best
fit. The bulk ($\sim75\%$) of the emission measure in the
0.5--8\,keV energy band, and therefore most of the intrinsic
X-ray emission, arises with a soft plasma component, with
$kT_{\rm soft}=0.9$\,keV (0.7--1.1\,keV); whereas a hard
plasma component, with $kT_{\rm hard}=4.2$\,keV
(3.0--6.5\,keV) contributes $\sim25\%$ of the total X-ray
emission. For this two-temperature model we find a column
density a bit larger than in model \#1, with $N_{\rm
  H}=4.1\times10^{22}$\,cm$^{-2}$
(3.5--$4.7\times10^{22}$\,cm$^{-2}$).  

One or two-temperature plasma models with the hydrogen
column density fixed to the lowest value found from
optical/IR observations of \vor~during outburst (e.g.,
$A_{\rm V}=$11\,mag from \cite{vacca04}, see discussion
below) lead to poor fits ($Q < 0.05$) and, hence, are not
acceptable. 

The K$\small \alpha$ fluorescent line from neutral to
low-ionized iron at 6.4\,keV was detected by
\cite*{imanishi01} with \cxo~from a class I protostar in the
$\rho$ Ophiuchi dark cloud during a bright flare, and was
interpreted as reflection emission from a face-on accretion
disk. We estimate the 90\% confidence limit on the strength
of a neutral 6.4\,keV iron line in the \vor~spectrum by
adding a Gaussian line of 12\,eV width at 6.4\,keV. This
will be important to constrain future 2D models involving
the heating of plasma and its interaction with nearby
circumstellar material. We retrieve exactly the best-fit 
parameters found for model \#2 (see Table~\ref{fit_parameters}), 
and the Gaussian line at 6.4\,keV
appears to improve the fit in the 6.7\,keV iron line region (see
Fig.~\ref{line}). However, we find that the total photon rate in
the line is $N_{\rm ph}=4.3\times10^{-7}$\,cm$^{-2}$\,s$^{-1}$
(0--8.7$\times10^{-7}$\,cm$^{-2}$\,s$^{-1}$) so, given this
range of uncertainty, we cannot exclude the possibility that
there is no 6.4\,keV iron line in the \xmm{} spectrum of \vor~.

Both the temperature and the column density found here with
the one-temperature plasma model are consistent with results
obtained from the same spectral modeling of \cxo~data
obtained on 2004 March 7. The large contribution from soft
plasma emission to the X-ray spectrum of \vor~revealed
by \xmm~data is reminiscent 
of that observed from \object{TW\,Hya}
(\cite{kastner02}). For TW\,Hya, the soft X-ray emission is
believed to be generated via mass accretion from its
circumstellar disk, which produces shocks onto the stellar
photosphere. However the hard plasma emission of \vor~above
1\,keV is too elevated to be explained in terms of accretion
shocks onto a low-mass star, and can be understood only in
the framework of plasma heating sustained by magnetic
reconnection events (e.g., \cite{kastner04}). 

\begin{figure*}[!t]
\begin{tabular}{@{}cc@{}}
\includegraphics[width=\columnwidth,bb=55 50 531 505, clip=true]{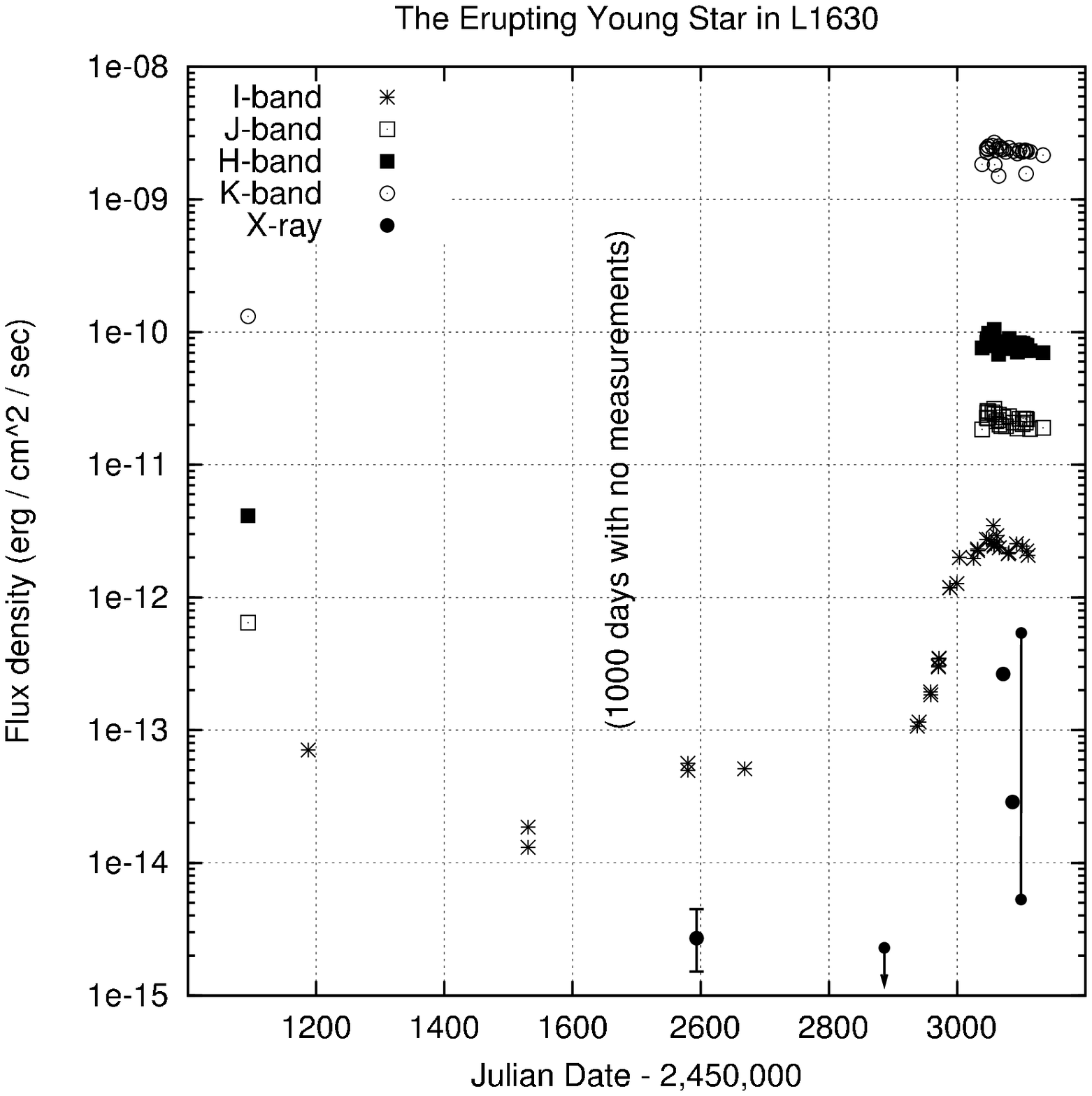} & \includegraphics[width=\columnwidth]{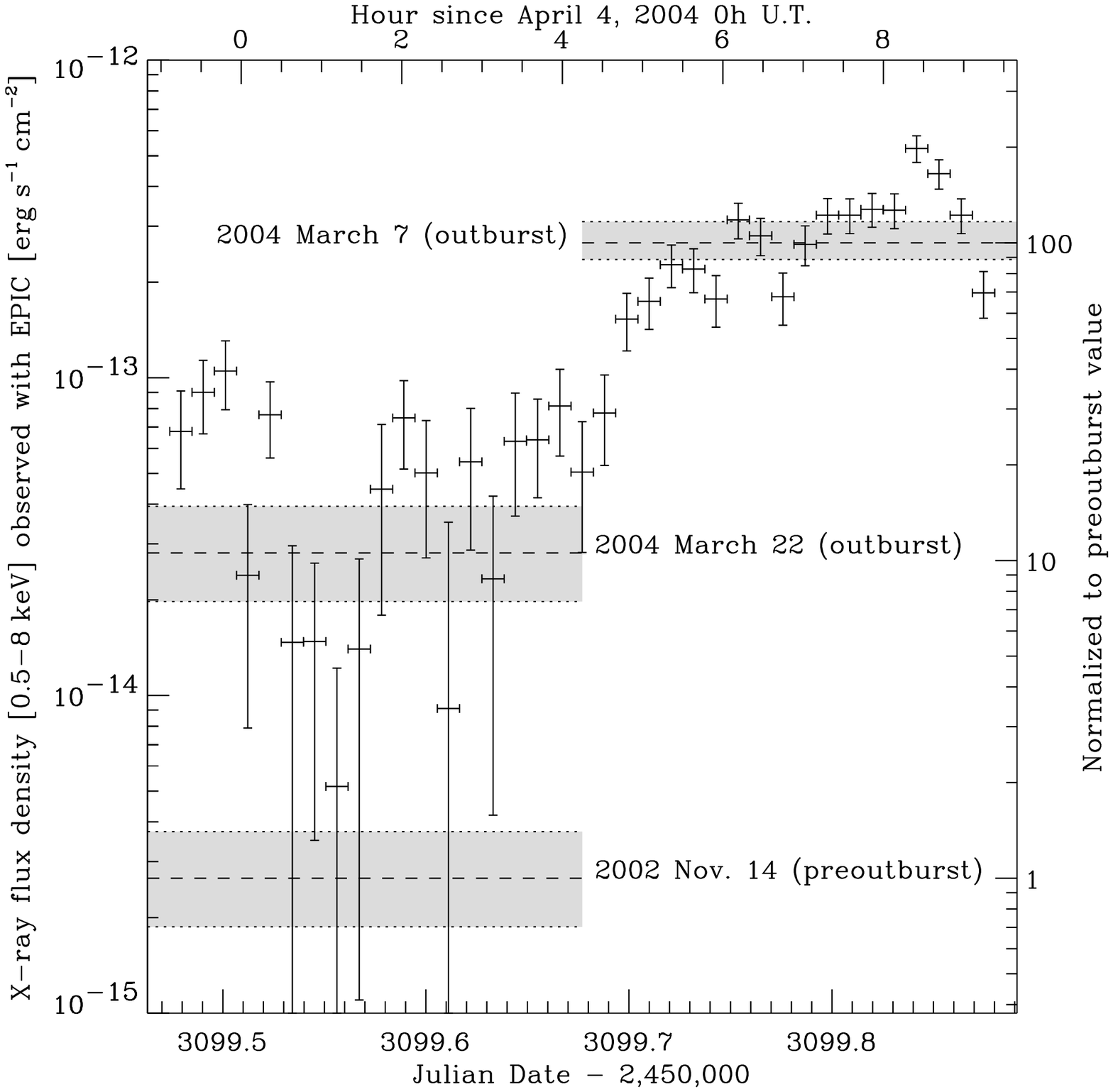} \\
\end{tabular}
\caption{Long term multiwavelength variation of the observed flux 
density of V1647\,Ori. The left hand panel (adapted from Kastner 
et al. 2004) shows the variation of the observed flux density of 
\vor~from the optical ($I$ band), near-IR ($J$, $H$, and $K$ bands), 
and X-ray (0.5--8\,keV energy band). The first three X-ray 
measurements are from \cxo~ (Kastner et al.\ 2004), where we have 
computed for the 2002 November 14 measurement the 90\% confidence 
level from observed counts using Gehrels (1986) statistics. The dotted 
arrow marks the X-ray upper limit at the 90\% confidence level 
computed from the \xmm~observation of HH24-26 on 2003 September 3 
(Ozawa et al., in preparation). The range of X-ray flux density that 
we report here with \xmm~is represented by a segment, detailed in 
the right hand panel. The right hand panel shows the X-ray flux 
density observed with \xmm, computed from the EPIC light curve 
(upper panel of Fig.~\ref{lc}) using a conversion factor from 
count rate to flux density derived from our best spectral fit 
(Fig.~\ref{spectra}). One sigma error bars are given for both 
\xmm~and \cxo~fluxes.}
\label{long_term}
\end{figure*}
\begin{figure*}[!t]
\vspace{0.5cm}
\begin{tabular}{@{}cc@{}}
\includegraphics[width=\columnwidth]{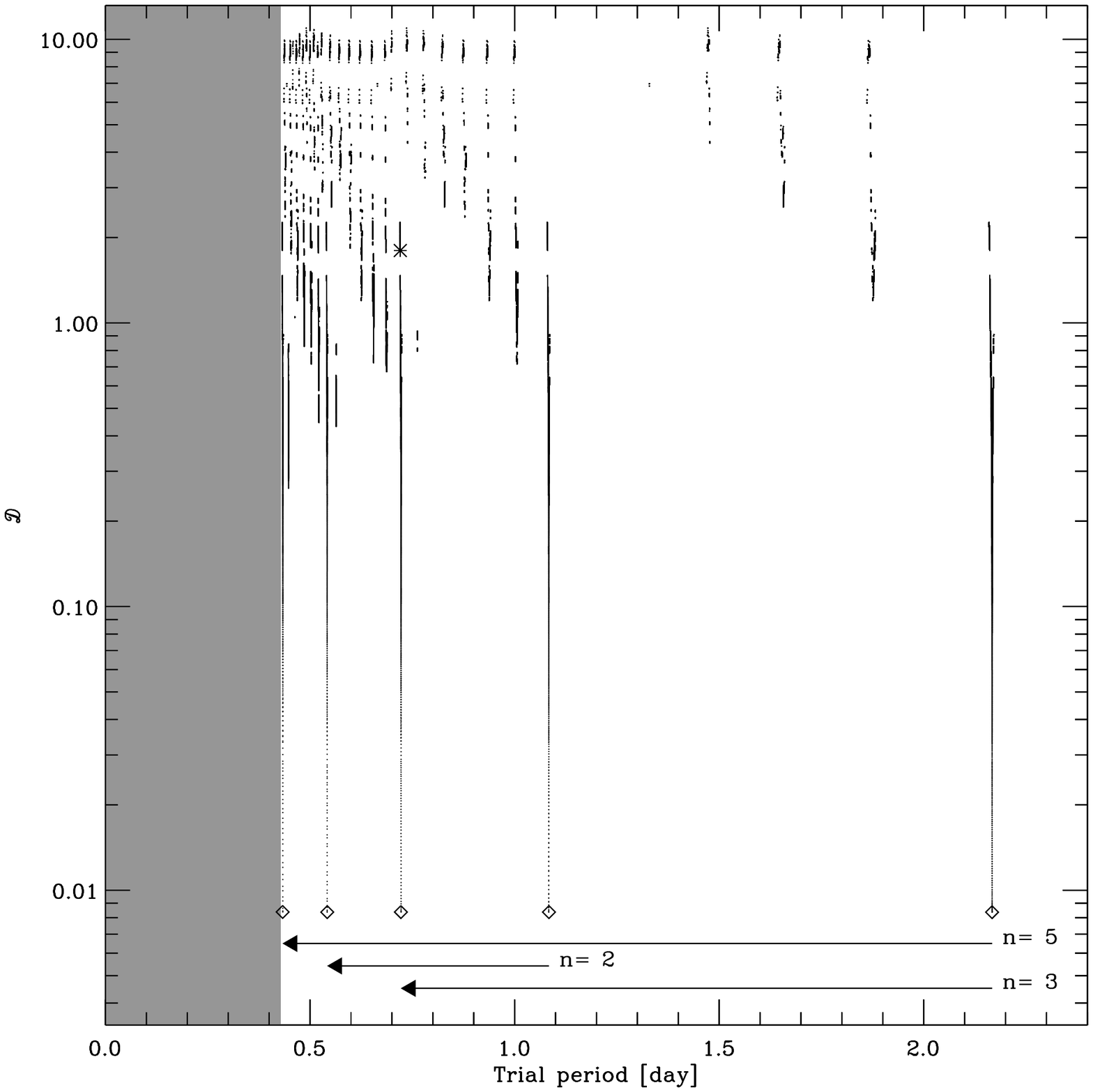} & \includegraphics[width=\columnwidth]{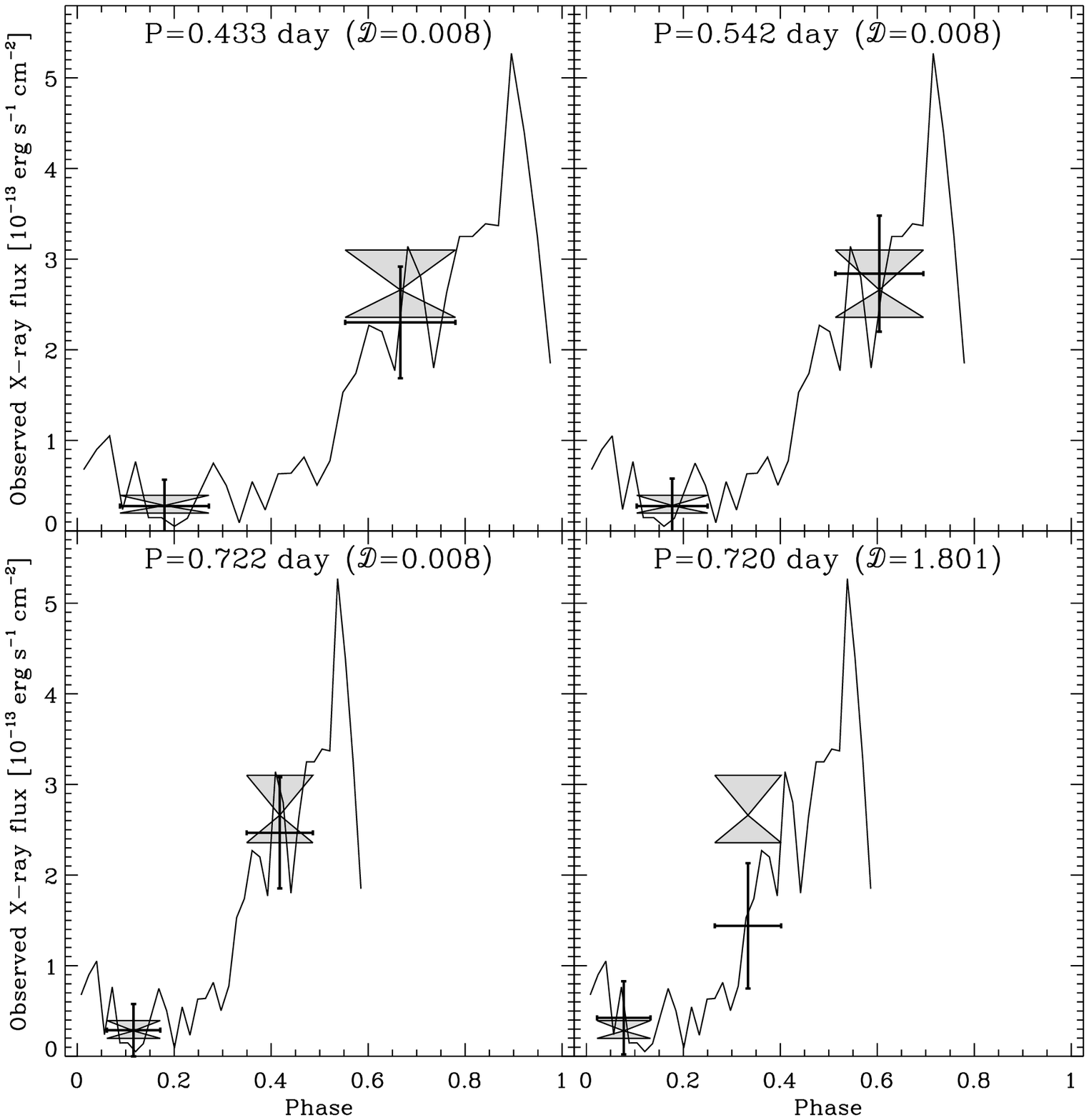} \\
\end{tabular}
\caption{Folded \xmm~and \cxo~X-ray fluxes of \vor. The left
  hand panel plots the function $\cal D$ versus trial
  period, from the value given by the length of the
  \xmm~observation to 2.3\,days with period step of
  $10^{-6}$\,s. This function is defined only for periods
  allowing the folding of the two \cxo~measurements with the
  \xmm~light curve (see text for details). Diamonds mark the
  best three shortest trial periods with their corresponding
  $n^{\rm th}$-harmonic. The right hand panel displays the
  resulting \xmm~(line) and \cxo~(grey hourglasses) ~flux
  measurements folded using these three periods. The error
  bars represent the equivalent \xmm~flux integrated during
  the \cxo~observing time. The last plot of the right hand
  panel shows for comparison a bad fit corresponding to the
  point marked by an asterisk in the left hand panel.} 
\label{period}
\end{figure*}

This \xmm~measurement of the hydrogen column density,
consistent with the previous \cxo~estimate, is now well
enough constrained to reveal a significant excess of
hydrogen column density compared to the value derived from
optical/IR observations. Combining $N_{\rm H}/A_{\rm
  J}=5.6\pm0.4\times 10^{21}$\,cm$^{-2}$\,mag$^{-1}$
(\cite{vuong03}), and $A_{\rm J}/A_{\rm
  V}=0.4008-0.1187\times(R_{\rm V}/3.1)^{-1}$
(\cite{cardelli89}), leads to $A_{\rm V}=(N_{\rm
  H}/10^{21})/[2.24-0.66\times(R_{\rm V}/3.1)^{-1}]$, where
the parameter $R_{\rm V}$ ranges from the average galactic
value of 3.1 to $\sim$5.5 for a dusty environment. Therefore
our measurement, $N_{\rm
  H}=3.5$--$4.7\times10^{22}$\,cm$^{-2}$, is equivalent to
$A_{\rm V}=$22.3--29.6\,mag and 18.8--25.0\,mag, for $R_{\rm
  V}$=3.1 and 5.5, respectively. The highest value found for
the visual extinction of \vor~in its outburst is $A_{\rm
  V}=$11\,mag (\cite{vacca04}), which is well below our 90\%
confidence interval. We stress that there remains a
discrepancy between our lower range estimate and this
highest visual extinction for any value of the parameter
$R_{\rm V}$. We conclude that there is an excess of hydrogen
column density in the local environment of V1647 Ori of at
least 1.5--2.6$\times 10^{22}$\,cm$^{-2}$, assuming the
visual extinction of \cite*{vacca04}, and $R_{\rm V} \sim
5.5$. The gas-to-dust ratio along the line-of-sight towards
\vor~is hence unusual, with more gas than dust towards this
object. This is consistent with the picture of the rise of a
wind/jet from the central embedded source, which would
increase the amount of gas close to central object where the
X-ray emission is believed to arise.

\section{Comparison of V1647\,Ori X-ray fluxes observed with Chandra and XMM-Newton}
\label{comparison}

We have updated Fig.~1 of \cite*{kastner04} by adding IR
measurements of \cite*{walter04}, which completes the
monitoring of \vor~through 2004 April, and \xmm~measurements
(see the left hand panel of
Fig.~\ref{long_term}). \xmm~observed serendipitously the
region of \vor~on 2003 September~3, with $\sim$50\,ks
exposure, by pointing toward the Herbig-Haro emission
complex HH24-26 located about $4\arcmin$ south (Ozawa et al., 
in preparation). \vor~was not detected. We convert this
non-detection (computed on a $\sim$25\,ks exposure without
flaring background) to a flux density upper limit assuming
an absorbed ($N_{\rm H}=5.7\times10^{22}$\,cm$^{-2}$)
one-temperature ($kT$=4.8\,keV) model with 
$Z$=0.3\,Z$_\odot$, i.e.\ the same model used by
\cite*{kastner04} to derive flux density from the
\cxo~detection of \vor~on 2002 November 14. The resulting
\xmm~upper limit is consistent with the low pre-outburst
activity level of \vor~observed by \cxo~on 2002 November 14
during a 55.9\,ks exposure. This demonstrates that the rise
of X-ray activity had not yet occured two months before the
optical/IR outburst and strengthens the apparent correlation
between the sharp increase in X-ray flux from pre-outburst
to outburst and the similar brightening seen in the
optical/IR, already noticed by \cite*{kastner04}. Using our
best-fit spectral model, we convert EPIC count rate in the
0.5--8\,keV energy range (see Fig.~\ref{lc}) to observed
X-ray flux density (note that the conversion factor is
essentially independent of the number of plasma
components). As shown in the right hand panel of
Fig.~\ref{long_term}, the X-ray flux observed with
\xmm~ranges from roughly the flux observed by \cxo~on 2004
March~22 (i.e.\ $\sim$10 times greater than the pre-outburst
X-ray flux) to a peak at a value two times greater than the
one caught by \cxo~on 2004 March~7 ($\sim$200 times greater
than the pre-outburst X-ray flux). The X-ray variability of
\vor~in outburst is hence also clearly {\sl enhanced\,}, as
noted by \cite*{kastner04}.   

The right hand panel of Fig.~\ref{long_term} reveals that 
the X-ray flux density of \vor~during the first half of the 
\xmm~observing period on 2004 April~4 is nearly identical 
to the flux density of the source during the epoch of the \cxo{} 
observations on 2004 March~22. Similarly, the enhanced flux density 
of \vor~during the second half of the \xmm~observing period is an 
excellent match to the X-ray brightness of the source during the 
epoch of the \cxo{} observations on 2004 March~7. The combined 
\cxo+\xmm~data set strongly suggests the possibility that the 
\vor~shows a periodic variation in X-ray brightness. We have 
investigated this possibility by folding the \cxo~flux 
measurements with those from \xmm. 

For this purpose, we define the following function:
\begin{displaymath}
{\cal D}(P) \equiv \sqrt{\sum_{i=1}^2 \frac{(F_{CXO,i}-\tilde{F}_{XMM,i,P})^2}{\sigma_{XMM,i,P}^2}}\, , 
\end{displaymath}
where $P$ is the trial period; $F_{CXO,i}$ is one of the two
\cxo~flux measurements; $\tilde{F}_{XMM,i,P}$ is the
\xmm~flux integrated over the \cxo~observational window
(depending of the trial period) using a linear interpolation
of the \xmm~data points; $\sigma_{XMM,i,P}$ is the standard
deviation of the \xmm~flux in the \cxo~observational
window. This function is defined only for periods allowing
the folding of the two \cxo~measurements with the \xmm~light
curve. ${\cal D}$ defines in the folded light curve the
quadratically averaged distance between \cxo~measurements
and the \xmm~light curve, in units of the standard deviation
of the \xmm~flux in the \cxo~observational window. We
computed numerically $\cal D$ versus trial periods, from the
value given by the length of the \xmm~observation to
2.3\,days, with period step of $10^{-6}$\,s. The left hand
panel of Fig.~\ref{period} shows the resulting value of
$\cal D$ for each trial period. The structure of the
$\cal D$ function thus obtained is non continuous and cyclical,
thereby appearing similar to an attractor. We consider the
periods providing the smallest value of $\cal D$ as the best period
candidates. The first three period candidates are: $P$=0.43,
0.54, and 0.72\,day, with ${\cal D}\sim 0.008$. The other
periods leading to a minimum value for $\cal D$ are $n^{\rm
  th}$-harmonics (where $n$ is an integer) of these first
periods. The right hand panel of Fig.~\ref{period} shows the
corresponding X-ray folded light curves obtained with these
period candidates. The last plot in Fig.~\ref{period}
illustrates a typical period-folding result when $\cal D$ is
greater than 1, corresponding to the asterisk point in the
left hand panel of Fig.~\ref{period}; clearly, in this case,
we obtain a poor fit. Because our combined \cxo~and
\xmm~dataset spans nearly 1 month, any folding of the
\cxo~flux measurements with periods much smaller than
$\sim$1 month is very sensitive to the assumed value of the
period. 

From both an observational and a physical point of view, the
0.72\,day period appears the most likely. Observationally,
with this period, roughly 60\% of the phase is covered by the
\xmm~observation, where during half of this
time interval the X-ray flux is close to one of the two levels measured with
\cxo. Of course if we assume that the light curve is indeed
symmetric about phase 0.5, then the X-ray flux is, during the
whole period, close to one of the two levels measured with
\cxo~precisely half the time. Therefore random short X-ray
observations would catch easily one of these two
characteristic level of activity, explaining naturally the
results obtained with \cxo. Physically, this period
corresponds to the time scale of Keplerian rotation at a
distance of 3.4 solar radius from a one solar mass star, or
equivalently to a distance of 1 (1.4) stellar radius for a
one solar mass star with an age of 0.5 (1.0) Myr
(\cite{palla99}).

\section{Discussion}
\label{discussion}

Until very recently, no X-ray satellites have had the opportunity 
to observe a PMS eruptive object at the beginning of its outburst. 
In this respect the recent \cxo~and \xmm~observations of \vor~offer 
new insight into a PMS accretion burst and, hence, into PMS accretion 
processes. During phases of normal accretion activity in T~Tauri 
stars ($\dot{M}_{acc}\sim10^{-7}$\,M$_\odot$\,yr$^{-1}$) the dynamic 
pressure from the disk accretion flow is balanced by the magnetic 
pressure of the stellar magnetosphere at the Alfv{\'e}n radius 
(\cite{ghosh78}; \cite{koenigl91}). The magnetic pressure inside this 
radius (close to the co-rotation radius) is strong enough to prevent 
the accreting material in the disk midplane from spiralling towards 
the star, but rather lifts it into free fall along the stellar field 
lines, leading to {\it magnetospheric accretion} (see review by \cite{shu00}). 

In contrast, during an accretion outburst, the ram pressure from 
the accretion disk flow is sufficient to push the inner edge of the 
accretion disk within the co-rotation radius, leading to strong 
magnetic interaction between the star magnetosphere and its inner 
accretion disk. This interaction area, called the {\it reconnection ring}, 
is then a possible source of hard and soft X-ray emission from magnetic 
reconnection events (\cite{shu97}). Strong reconnection events will 
also occur when differential rotation between the inner disk edge 
and stellar magnetospshere cause fields lines to twist, strengthen 
and expand (\cite{lovelace95}; \cite{goodson97}; \cite{matt02}). 
In these cases reconnection drives outflows occuring periodically 
above and below the disk (\cite{goodson99a}; \cite{goodson99b}).

In the most extreme case, the accretion flow is even sufficient to 
crush the stellar magnetic field against the stellar photosphere, 
leading to the formation of a {\it boundary layer} at the inner edge 
of the accretion disk (\cite{hartmann98}), which may quench the 
star-disk X-ray emission.

The previous shorter (5\,ks exposures) observations of \vor~obtained 
with \cxo~have shown on 2004 March~7 during the outburst an enhanced 
X-ray flux compared to the pre-outburst level measured by \cxo~on 2002 
November 14; and a drop of the X-ray flux on 2004 March~22, which we 
previously interpreted as the possible onset of a quenching X-ray 
emission phase, or the triggering of a phase of strong variability 
in both X-ray luminosity and temperature (\cite{kastner04}). The longer 
$\sim38$\,ks observation with \xmm~of \vor~on 2004 April~7, presented 
here, shows that there is indeed as yet no quenching X-ray emission phase, 
but there is clearly enhanced X-ray variability. 

The detailed X-ray spectrum obtained with \xmm~improves of
our knowledge of the X-ray properties during the outburst
phase, showing that $\sim75\%$ of the intrinsic X-ray
emission in the 0.5--8\,keV energy band may be explained by
an accretion shock onto the photosphere of this low-mass
star, whereas $\sim25\%$ of the total X-ray emission comes
from a plasma heated by a magnetic reconnection
mecanism. The dramatic increase of accretion rate, producing
the optical/IR outburst, may also explain the observed rise
of X-ray emission from \vor: first by increasing the
quantity of gas falling onto the stellar surface, which
radiates soft X-rays at the accretion shock; second by
pushing the inner accretion disk boundary inside the
corotation radius, which would increase the contribution of
the reconnection ring to the soft/hard X-ray emission. 

Our quantile analysis has shown that large increase of the
count rate is not associated with any large temperature
variations as for typical X-ray flares from YSOs, suggesting
that we are likely also observing variations of the emission
measure. The variation of the emission measure can be
produced by a variation of the (observed) plasma volume
and/or a variation of the plasma electronic density. We
cannot totally exclude that we observed an unusual X-ray
flare with a strong variation with time of the plasma
electronic density in the flaring loop. However, the
comparison of our \xmm~light curve with the
\cxo~measurements has suggested that the X-ray flux of
\vor~could be periodic, with a likely period of 0.72\,day --
corresponding to the Keplerian rotation of close material in
the close vicinity of the star. Therefore, the enhanced
X-ray variability of \vor~is more likely explained by a
variation of the observed plasma volume by rotational
modulation. In this scenario, the inner part of the
accretion disk eclipses periodically a fraction of the
plasma volume close to the stellar surface, producing a
dimming of the observed X-ray flux by a factor of about
20. Rotational modulation by accretion funnels has been
proposed to explain optical eclipses observed in the
classical T~Tauri star AA\,Tau, where the accretion disk is
thought to be warped by the magnetospheric magnetic dipole
inclined respectively to the star's spin (\cite{bouvier99}).  

We stress that the optical/IR flux of \vor~is dominated by 
the outburst luminosity of its inner accretion disk and scattered 
light from the surrounding nebulosity, therefore any optical/IR 
modulations of the photospheric flux are difficult if not impossible 
to observe. Therefore X-rays, likely produced only close to the 
central star, offer a unique observational window to probe the 
star-disk interaction area during this PMS evolutionary phase 
in which the accretion rate, and hence the source luminosity, 
are dramatically elevated above normal.

Longer X-ray observations with \xmm~and \cxo~will be needed to 
confirm both the candidate period, and to validate the scenario 
proposed here to explain the enhanced X-ray variability.

\begin{acknowledgements}
We would like to thank the Target of Opportunity panel of \xmm~and 
the \xmm's project scientist Norbert Schartel, who gave us the 
opportunity to observe \vor{} during its outburst. We thanks the 
anonymous referee for his comments and suggestions. H.~O. acknowledges 
the support of the Conseil National des Astronomes et Physiciens. 
Based on observations obtained with the \xmm, an ESA science mission 
with instruments and contributions directly funded by ESA member states 
and the USA (NASA). We used archival acquisition image made with ESO 
Very Large Telescope at Paranal Observatory under programme ID 272.C-5045.
\end{acknowledgements}


\begin{thebibliography}{}
\bibitem[\protect\astroncite{{\'A}brah{\'a}m et al.}{2004}]{abraham04} 
{\'A}brah{\'a}m, P., K{\'o}sp{\'a}l, {\'A}., Csizmadia, S., et al.\ 2004,  \aap, 419, L39
\bibitem[\protect\astroncite{Andrews et al.}{2004}]{andrews04} 
Andrews, S.~M., Rothberg, B., \& Simon, T.\ 2004,  \apjl,  610, L45
\bibitem[\protect\astroncite{Anthony-Twarog}{1982}]{anthony82} 
Anthony-Twarog, B.~J.\ 1982,  \aj,  87, 1213
\bibitem[\protect\astroncite{Aspin \& Reipurth}{2004}]{aspin04} 
Aspin, C.~\& Reipurth, B.\ 2004,  \iaucirc,  8396, 3
\bibitem[\protect\astroncite{Bouvier et al.}{1999}]{bouvier99} 
Bouvier, J., Chelli, A., Allain, S., et al.\ 1999,  \aap,  349, 619
\bibitem[\protect\astroncite{Brice{\~ n}o et al.}{2004}]{briceno04} 
Brice{\~ n}o, C., Vivas, A.~K., Hern{\'a}ndez, J., et al.\ 2004, \apjl,  606, L123
\bibitem[\protect\astroncite{Cardelli et al.}{1989}]{cardelli89} 
Cardelli, J.~A., Clayton, G.~C., \& Mathis, J.~S.\ 1989,  \apj,  345, 245
\bibitem[\protect\astroncite{Clark}{1991}]{clark91} 
Clark, F.~O.\ 1991,  \apjs,  75, 611
\bibitem[\protect\astroncite{Eisl{\"o}ffel \& Mundt}{1997}]{eisloffel97} 
Eisl{\"o}ffel, J.~\& Mundt, R.\ 1997,  \aj,  114, 280
\bibitem[\protect\astroncite{Gehrels}{1986}]{gehrels86} 
Gehrels, N.\ 1986, \apj, 303, 336
\bibitem[\protect\astroncite{Ghosh \& Lamb}{1978}]{ghosh78} 
Ghosh, P.~\& Lamb, F.~K.\ 1978,  \apjl,  223, L83
\bibitem[\protect\astroncite{Goodson et al.}{1997}]{goodson97} 
Goodson, A.~P., Winglee, R.~M., \& Boehm, K.\ 1997,  \apj,  489, 199
\bibitem[\protect\astroncite{Goodson \& Winglee}{1999}]{goodson99a} 
Goodson, A.~P.~\& Winglee, R.~M.\ 1999,  \apj,  524, 159
\bibitem[\protect\astroncite{Goodson et al.}{1999}]{goodson99b} 
Goodson, A.~P., B{\" o}hm, K., \& Winglee, R.~M.\ 1999,  \apj,  524, 142
\bibitem[\protect\astroncite{Hartmann \& Kenyon}{1996}]{hartmann96} 
Hartmann, L.~\& Kenyon, S.~J.\ 1996,  \araa,  34, 207
\bibitem[\protect\astroncite{Hartmann}{1998}]{hartmann98} 
Hartmann, L.\ 1998, {\it Accretion processes in star formation} (Cambridge : Cambridge University Press)
\bibitem[\protect\astroncite{Herbig}{1966}]{herbig66} 
Herbig, G.~H.\ 1966,  Vistas in Astronomy,  8, 109
\bibitem[\protect\astroncite{Herbig}{1977}]{herbig77} 
Herbig, G.~H.\ 1977,  \apj,  217, 693
\bibitem[\protect\astroncite{Herbig et al.}{2001}]{herbig01} 
Herbig, G.~H., Aspin, C., Gilmore, A.~C., Imhoff, C.~L., \& Jones, A.~F.\ 2001,  \pasp,  113, 1547
\bibitem[\protect\astroncite{Hong et al.}{2004}]{hong04} 
Hong, J., Schlegel, E.~M., \& Grindlay, J.~E. 2004, \apj, 614, 508
\bibitem[\protect\astroncite{Imanishi et al.}{2001}]{imanishi01} 
Imanishi, K., Koyama, K., \& Tsuboi, Y.\ 2001,  \apj,  557, 747
\bibitem[\protect\astroncite{Imanishi et al.}{2003}]{imanishi03} 
Imanishi, K., Nakajima, H., Tsujimoto, M., Koyama, K., \& Tsuboi, Y.\ 2003, \pasj,  55, 653
\bibitem[\protect\astroncite{Jansen et al.}{2001}]{jansen01} 
Jansen, F., Lumb, D., Altieri, B., et al.\ 2001,  \aap,  365, L1
\bibitem[\protect\astroncite{Kaastra et al.}{1996}]{kaastra96} 
Kaastra, J.~S., Mewe, R., \& Nieuwenhuijzen, H.\ 1996,  in {\it UV and X-ray Spectroscopy of Astrophysical and Laboratory Plasmas}, ed. K.~Yamashita \& T.~Watanabe (Tokyo : Univ. Acad. Press), p.~411
\bibitem[\protect\astroncite{Kastner et al.}{2002}]{kastner02} 
Kastner, J.~H., Huenemoerder, D.~P., Schulz, N.~S., Canizares, C.~R., \& Weintraub, D.~A.\ 2002,  \apj,  567, 434
\bibitem[\protect\astroncite{Kastner et al.}{2004}]{kastner04} 
Kastner, J.~H., Richmond, M., Grosso, N., et al.\ 2004,  \nat,  430, 429
\bibitem[\protect\astroncite{Kenyon \& Hartmann}{1991}]{kenyon91} 
Kenyon, S.~J.~\& Hartmann, L.~W.\ 1991,  \apj,  383, 664
\bibitem[\protect\astroncite{Koenigl}{1991}]{koenigl91} 
Koenigl, A.\ 1991,  \apjl,  370, L39
\bibitem[\protect\astroncite{Kun et al.}{2004}]{kun04} 
Kun,  M., Acosta-Pulido, J.~A., Mooret, A., et al.\ 2004, A\&A, submitted [astro-ph/0408432]
\bibitem[\protect\astroncite{Lis et al.}{1999}]{lis99} 
Lis, D.~C., Menten, K.~M., \& Zylka, R.\ 1999,  \apj,  527, 856
\bibitem[\protect\astroncite{Lovelace et al.}{1995}]{lovelace95} 
Lovelace, R.~V.~E., Romanova, M.~M., \& Bisnovatyi-Kogan, G.~S.\ 1995, \mnras,  275, 244
\bibitem[\protect\astroncite{Lupton et al.}{2004}]{lupton04} 
Lupton, R., Blanton, M.~R., Fekete, G., et al.\ 2004,  \pasp,  116, 133
\bibitem[\protect\astroncite{Mallas \& Kreimer}{1978}]{mallas78} 
Mallas, J.~H., \& Kreimer, E.\ 1978, Sky Publication Co., Cambridge, Mass. 
\bibitem[\protect\astroncite{Mason et al.}{2001}]{mason01} 
Mason, K.~O., Breeveld, A., Much, R., et al.\ 2001,  \aap,  365, L36
\bibitem[\protect\astroncite{Matt et al.}{2002}]{matt02} 
Matt, S., Goodson, A.~P., Winglee, R.~M., \& B{\" o}hm, K.\ 2002,  \apj,  574, 232
\bibitem[\protect\astroncite{McGehee et al.}{2004}]{mcgehee04} 
McGehee,  P.~M., Smith, J.~A., Hende, A.~A., et al.\ 2004, \apj, 616, 1058
\bibitem[\protect\astroncite{McLaughlin}{1946}]{mcLaughlin46} 
McLaughlin, D.~B.\ 1946,  \aj,  52, 109
\bibitem[\protect\astroncite{McNeil}{2004}]{mcneil04} 
McNeil, J.~W.\ 2004,  \iaucirc,  8284, 1
\bibitem[\protect\astroncite{Mitchell et al.}{2001}]{mitchell01} 
Mitchell, G.~F., Johnstone, D., Moriarty-Schieven, G., Fich, M., \& Tothill, N.~F.~H.\ 2001,  \apj,  556, 215
\bibitem[\protect\astroncite{Morrison \& McCammon}{1983}]{morrison83} 
Morrison, R.~\& McCammon, D.\ 1983,  \apj,  270, 119
\bibitem[\protect\astroncite{Muzerolle et al.}{2005}]{muzerolle05} 
Muzerolle, J., Megeath, S.~T., Flaherty, K.~M., et al.\ 2005,  \apjl,  620, L107
\bibitem[\protect\astroncite{Palla \& Stahler}{1999}]{palla99} 
Palla, F.~\& Stahler, S.~W.\ 1999,  \apj,  525, 772
\bibitem[\protect\astroncite{Reipurth \& Aspin}{2004}]{reipurth04} 
Reipurth, B.~\& Aspin, C.\ 2004,  \apjl,  606, L119
\bibitem[\protect\astroncite{Rettig et al.}{2005}]{rettig05} 
Rettig, T.~W., Brittain, S.~D., Gibb, E.~L., Simon, T., Kulesa, \& C.~A.\ 2004, \apj, in~press
\bibitem[\protect\astroncite{Samus}{2004}]{samus04} 
Samus, N.~N.\ 2004, \iaucirc,  8354, 1
\bibitem[\protect\astroncite{Simon et al.}{2004}]{simon04} 
Simon, T., Andrews, S.~M., Rayner, J.~T., \& Drake, S.~A.\ 2004,  \apj,  611, 940
\bibitem[\protect\astroncite{Shu et al.}{1997}]{shu97} 
Shu, F.~H., Shang, H., Glassgold, A.~E., \& Lee, T.\ 1997, Science, 277, 1475 
\bibitem[\protect\astroncite{Shu et al.}{2000}]{shu00} 
Shu, F.~H., Najita, J.~R., Shang, H., \& Li, Z.-Y.\ 2000, Protostars \& Planets IV,  789
\bibitem[\protect\astroncite{Str{\" u}der et al.}{2001}]{strueder01} 
Str{\" u}der, L., Briel, U., Dennerl, K., et al.\ 2001, \aap, 365, L18 
\bibitem[\protect\astroncite{Tsuboi et al.}{1998}]{tsuboi98} 
Tsuboi, Y., Koyama, K., Murakami, H., et al.\ 1998,  \apj,  503, 894
\bibitem[\protect\astroncite{Turner et al.}{2001}]{turner01} 
Turner, M.~J.~L., Abbey, A., Arnaud, M., et al.\ 2001, \aap, 365, L27
\bibitem[\protect\astroncite{Vacca et al.}{2004}]{vacca04} 
Vacca, W.~D., Cushing, M.~C., \& Simon, T.\ 2004, \apjl,  609, L29
\bibitem[\protect\astroncite{Vuong et al.}{2003}]{vuong03} 
Vuong, M.~H., Montmerle, T., Grosso, et al.\ 2003,  \aap,  408, 581
\bibitem[\protect\astroncite{Walter et al.}{2004}]{walter04} 
Walter, F.~M., Stringfellow,  G.~S., Sherry, W.~H., Pollatou, A.~F. 2004, \aj, 128, 1872
\end{thebibliography}
\end{document}